\documentclass[11pt,twoside]{article}

\usepackage{asp2006_hotcool}
\usepackage{epsf}
\usepackage{lscape}

\usepackage{graphicx}

\begin{document}
\title{Constraining the evolution of red supergiants with the
integrated light of star clusters.}   
\author{Ariane Lan\c{c}on \& Morgan Fouesneau}   
\affil{Observatoire Astronomique (UMR 7550), Universit\'e de Strasbourg
\& CNRS, 11 rue de l'Universit\'e, 67000 Strasbourg, France}   

\begin{abstract} 
The integrated properties of young star clusters are subject to
large cluster-to-cluster variations because they depend directly on
small numbers of bright stars. At ages at which red supergiants are
expected to exist, these luminous but rare stars can be blamed
for most of the variations. If unresolved clusters are to be used 
to constrain red supergiant models, methods must be developed that 
take these stochastic fluctuations into account. We discuss
prospects and open issues in this field, based on recent
work on high mass clusters in M\,82, and on first experiments
towards a Bayesian study of cluster populations.
\end{abstract}

\section{Introduction}

Stellar astrophysics has a long tradition of using star clusters to test 
and improve evolutionary models\,: with small internal spreads in age 
and metallicity, clusters do indeed provide an extraordinary testbed. 
It is natural to consider using young clusters when studying
the evolution and the properties of massive stars. 
In this contribution, we focus on the red supergiant
phase of the evolution of relatively massive stars (typically
7--25\,M$_{\odot}$), i.e. cluster ages of about 8 to 60\,Myr.
We also limit the discussion to clusters that can not be resolved into
individual stars.

Red supergiants (RSGs) are intrinsically rare objects, as a combined result of
the stellar Initial Mass Function (IMF) and of the short 
duration of the relevant evolutionary phase. At cluster ages of
10 to 30\,Myr, their expectation numbers are of order one for total
cluster masses of order $10^4$\,M$_{\odot}$ for a standard IMF with
a lower mass cut-off of 0.1\,M$_{\odot}$ (see Tab.\,3 of 
Lan\c{c}on et al. 2008). Thus, individual clusters of 10$^4$\,M$_{\odot}$ 
might host zero, one, two or a few red supergiants, which 
translates into multimodal probability distributions for colours such
as $(V-K)$. It takes clusters of several 10$^6$\,M$_{\odot}$ to 
reduce random variations in the K band flux and in $(V-K)$ to
less than about 5\,\%, in which case flux and colour distributions are also
single peaked and gaussian approximations become tolerable. 
Interesting work on flux and colour distributions
at various cluster ages include Barbaro \& Bertelli (1977), 
Girardi \& Bica (1993), Lan\c{c}on \& Mouhcine (2000), Bruzual (2001),
Cervi\~no et al.  (2002, 2006).

The effects of changes in stellar model assumptions can be tested
by direct comparisons with the integrated properties of 
individual star clusters only
if they are larger  than the spread resulting from the
above-mentioned small numbers of luminous stars, the so-called 
``stochastic fluctuations". 

As discussed in several contributions to this meeting, models relevant to
the red supergiant phase of evolution remain particularly uncertain. 
This is seen when comparing theoretical stellar spectra 
with empirical spectra of luminous cool stars
(Lan\c{c}on et al. 2007), or when using different sets of
evolutionary tracks in population synthesis calculations. 
For instance, Gonz\'alez Delgado et al. (2005) compare synthetic
spectra of Single Stellar Populations (SSP) based on the 
tracks of the Geneva group and the Padova group, respectively
(Schaller et al. 1992, Girardi et al. 2000); 
and V\'azquez et al. (2007) consider tracks with and
without inital rotation (Meynet \& Maeder 2005).
The effects of currently acceptable changes in model assumptions are 
large enough that we can expect constraints from the integrated
light of star clusters despite the stochastic fluctuations.
Possibilities have not yet been exploited fully, because 
replacing evolutionary tracks or stellar spectral libraries
in population synthesis models is not as trivial as one might wish,
and because probabilistic methods for tackling low mass clusters
are still under development. First results highlight prospects 
and difficulties.

\section{Individual massive star clusters}

Massive young star clusters are found in actively star forming galaxies,
where the total number of clusters formed is large enough for the
upper end of the cluster mass function to be populated. An example
is M\,82. Five clusters more massive than 5\,$10^5$\,M$_{\odot}$
in M\,82 were studied by Lan\c{c}on et al. (2008; L08 hereafter), based on
extended near-IR spectra ($0.8-2.4\,\mu$m; $\lambda/\delta \lambda \sim 1000$)
and pre-existing optical data. All these clusters 
have near-IR spectral signatures
that immediately suggest a strong contribution from red supergiants. 
This is a common and expected property of starburst populations.

Using a population synthesis tool based on {\sc Pegase} (Fioc \& 
Rocca-Vol\-me\-ran\-ge 1997), with the evolutionary
tracks of Bressan et al. (1993) and a new
empirical near-IR library, L08 showed that it was possible to obtain
very good fits to the near-IR spectra of the star clusters in M82. The 
derived near-IR ages lie between 9 and 40\,Myr, i.e. in the age range where
the contributions of luminous red supergiants (class Ia and Iab)
are strongest in the models. 

However, the near-IR ages do not always agree with the best 
optical ages in the previous literature. The optical studies most 
frequently use the spectral region around the Balmer jump, a 
wavelength range at which red supergiants contribute very little, 
but which is affected by light from stars on the blue loops
of evolutionary tracks. At the ages derived from optical studies for
three clusters, the red supergiant features in the model spectra of 
L08 tend to be too weak.

What does this tell us about red supergiants models?

The spectra of massive clusters in M82 favour models that 
give luminous red supergiants a strong weight in the near-IR
light. This can be achieved in various ways, among which: \\
-- longer lifetimes in the red phases of evolution; \\
-- higher assigned gravities for the library stars of class Ia and Iab
(with higher assigned gravities, these spectra contribute to the 
emission of SSPs over a wider range of ages); \\
-- changes in the adopted bolometric corrections.

In order to finalise this study and determine which set of
evolutionary models and model atmosphere inputs are best able
to reproduce reality in M\,82, it will be necessary to obtain
even better spectra. At spectral resolutions of about 1000, it is critical 
to use optical and near-IR data jointly, otherwise the problem
is underconstrained. Observational apertures must be very precisely matched.  
Indeed, high resolution images of the clusters in M\,82 show that there
is significant substucture on the scale of a single cluster, with 
extinction lanes covering parts of them, and contaminating nebular or 
stellar sources on very close lines of sight (e.g. Bastian et al. 2007).
It is hoped that instruments such as the upcoming ESO/VLT/Xshooter
will be available for observational programmes of this type in
Southern Hemisphere galaxies.

Of course, one must keep in mind that the most compact massive star 
clusters might not be the best templates for stellar evolution as a whole. 
With thousands of stars within each cubic parsec, 
their red supergiants could be affected by neighbourhood effects; 
interacting binaries and multiple systems may be more frequent than elsewhere
(see the discussion focused on AGB stars by Gallagher \& Smith, 2007,
and references therein). 
Also, massive clusters might not all contain simple coeval stellar populations.
Self-enrichment could occur over timescales that are under debate.
Globular clusters, for instance, display abundance patterns that are
attributed to short-lived massive stars by some authors (Decressin et al. 2007),
to longer-lived intermediate mass stars by others (D'Ercole et al. 2008).
Finally, some clusters may be the results of mergers. The analysis of
high quality, extended spectra of massive clusters with standard models 
will show whether additional physical phenomena such as these need to
be taken into account. We suspect that the effects of non-instantaneous
star formation and of binarity on the integrated properties may be 
testable despite the stochastic fluctuations.

\section{Clusters of more common total masses}

\begin{figure}
\includegraphics[clip=,width=0.55\textwidth]{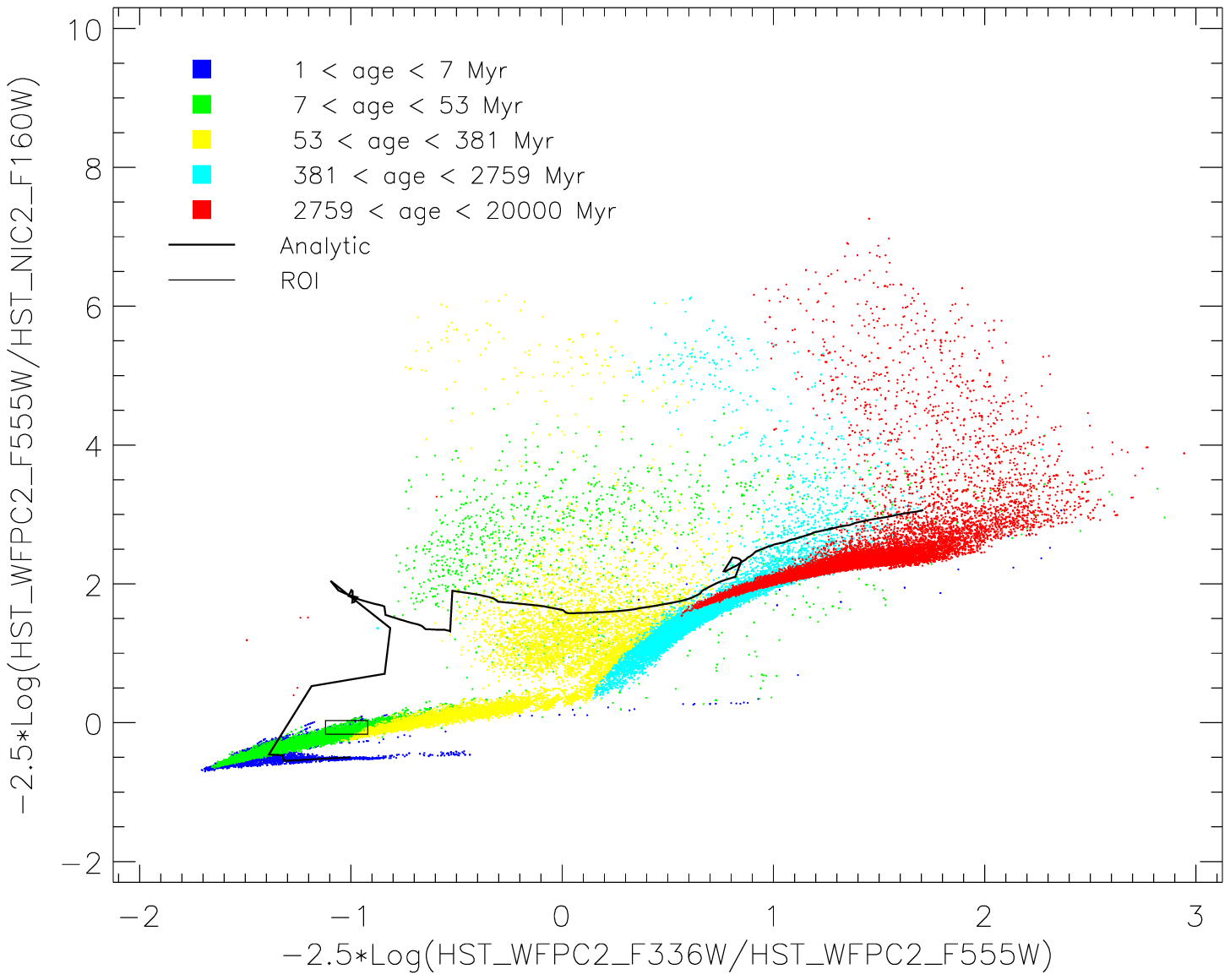}
\hspace{-0.078\textwidth}
\includegraphics[clip=,width=0.55\textwidth]{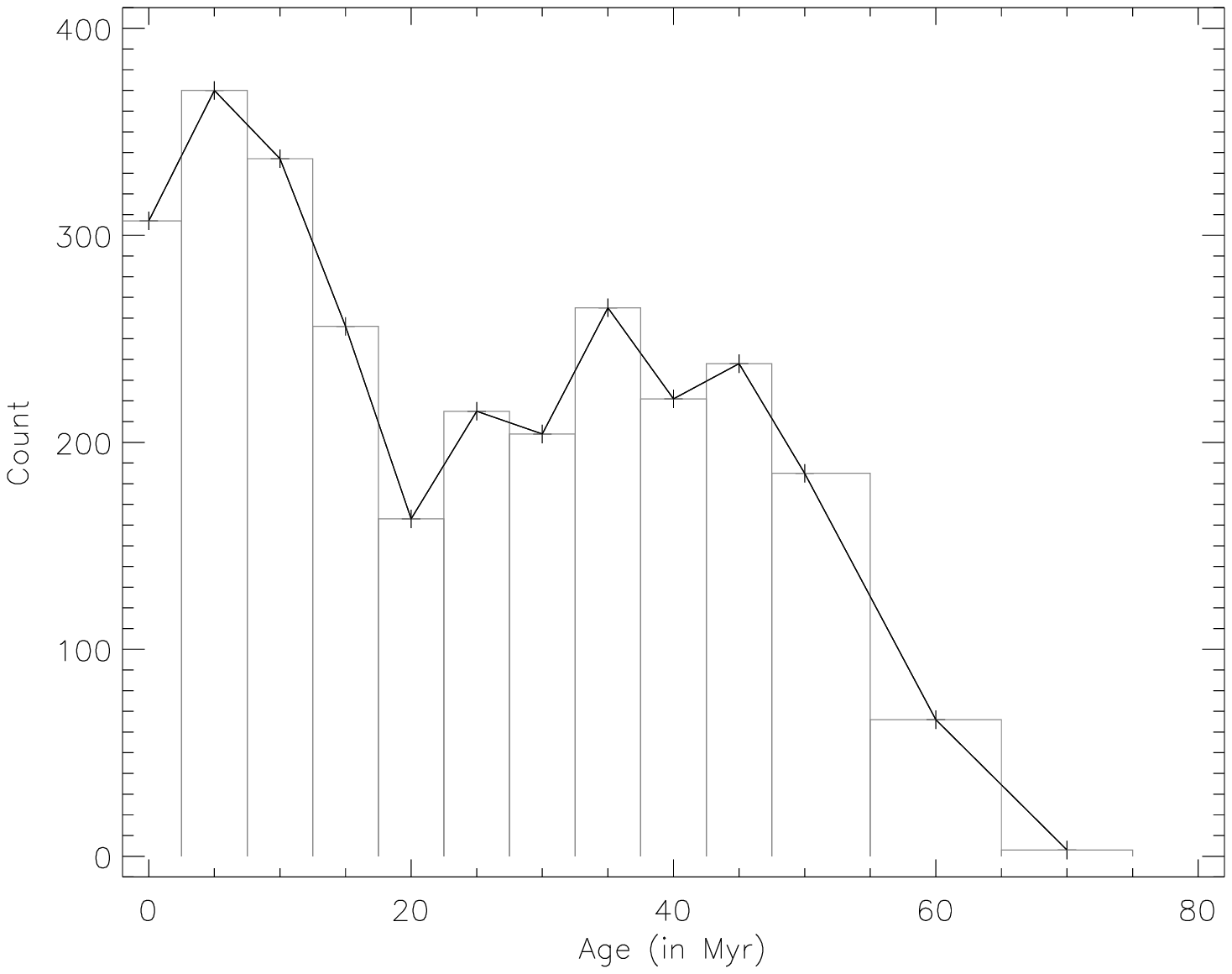}\\
\includegraphics[clip=,width=0.55\textwidth]{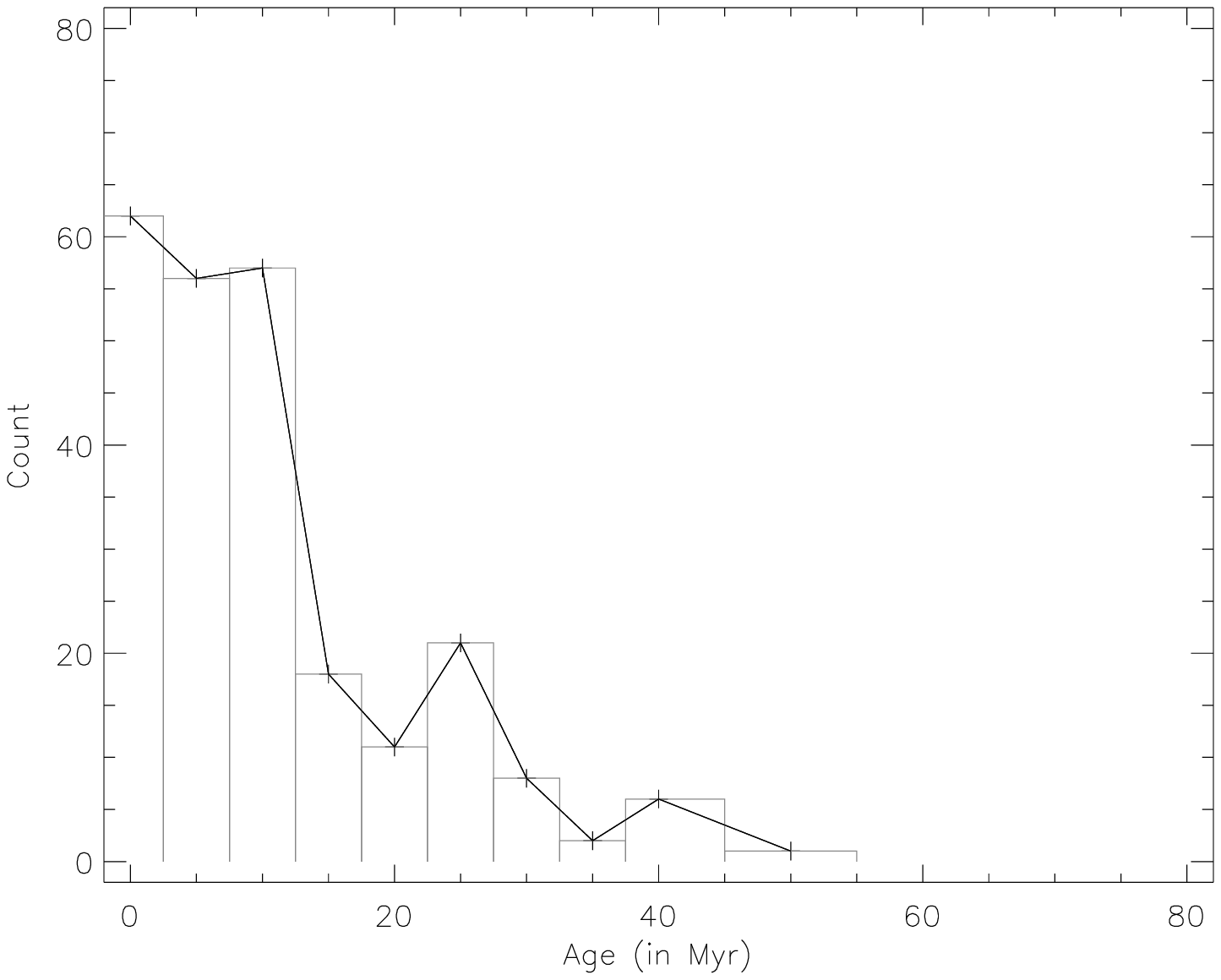}
\hspace{-0.078\textwidth}
\includegraphics[clip=,width=0.55\textwidth]{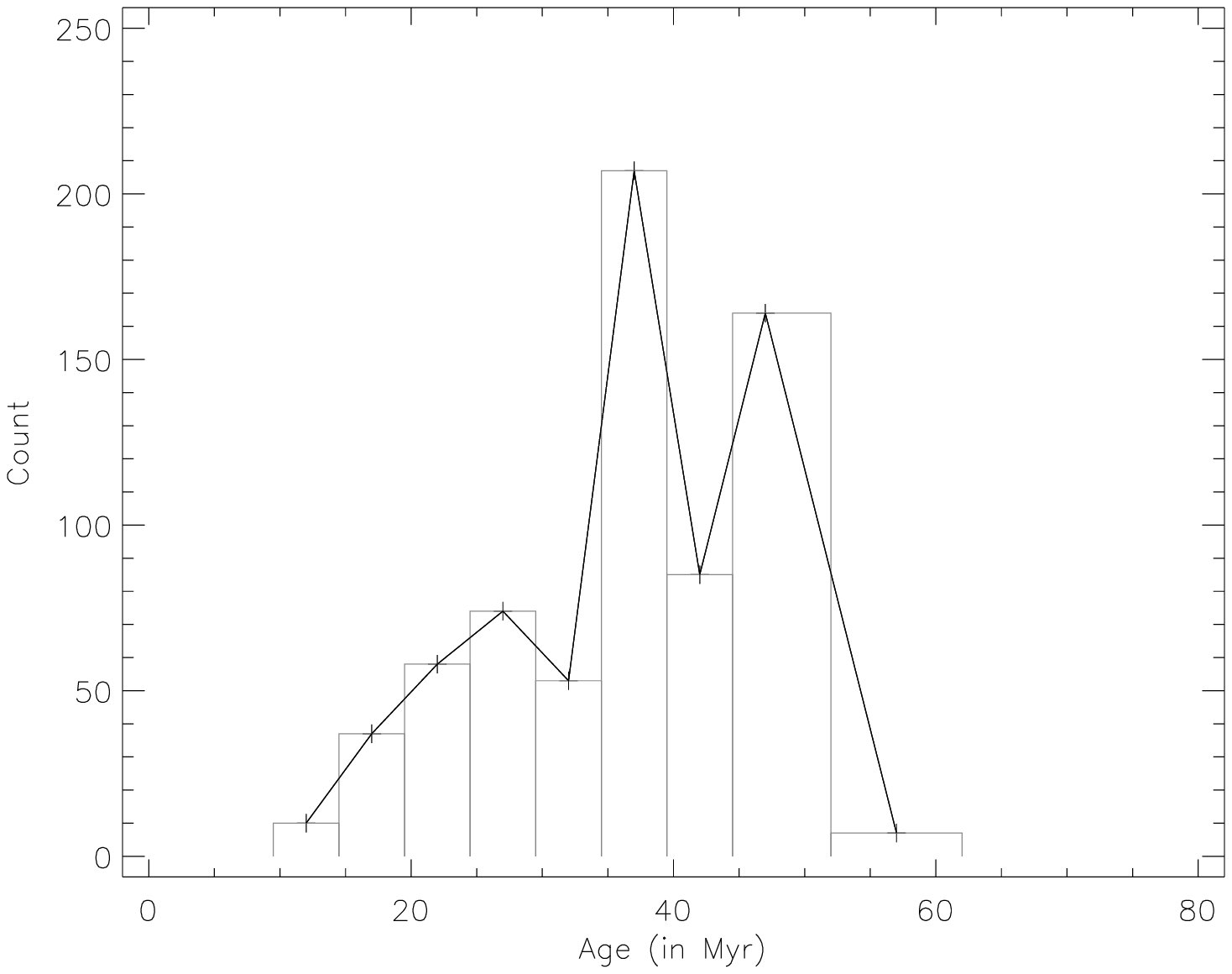}\\
\vspace*{-0.8\textwidth}

~\hspace{0.65\textwidth}\parbox[t]{0.3\textwidth}{\footnotesize
  (U-V) \& (V-H) : \\
  2830 clusters of $10^3$ stars}
\vspace{0.35\textwidth}

~\hspace{0.10\textwidth}\parbox[t]{0.35\textwidth}{\footnotesize
  (U-V), (V-H), (F255W-V) \& V : \\
  242 clusters of $10^3$ stars}
\hspace{0.15\textwidth}
  \parbox[t]{0.35\textwidth}{\footnotesize
  (U-V) \& (V-H) : \\
  695 clusters of $6\,10^3$ stars}
\vspace{0.3\textwidth}
  
\caption[]{{\bf Top left:}  HST colors of 69000 star clusters each containing
10$^3$ stars. The colours shown combine fluxes in F336W (U) and 
F555W (V) of WFPC2, and F160W (H) of NICMOS2.
Ages are distributed uniformly on a logarithmic scale 
between 1\,Myr and 20\,Gyr. The solid line shows the evolution
of the average colours (average over an infinite number
of clusters of the same age) with age. Solar metallicity is assumed. 
The box near coordinates [-1;0] surrounds the locus of one particular
8\,Myr old cluster, and illustrates 0.1\,magnitude error bars.
{\bf Top right:} Age distribution of the clusters with (U-V) and 
(V-H) colours within the above-mentioned error box (2830 clusters
out of 69000).
{\bf Bottom left:} Age distribution of the clusters found when 
the selection includes one more ultraviolet band and the absolute
V magnitude (242 clusters out of 69000).
{\bf Bottom right:} Age distribution obtained with a selection as in
the upper right pannel, but using a database of clusters that all
contain 6\,10$^3$ stars instead of $10^3$ (695 clusters out of 69000).\\
{\sc [nb: top left pannel appears properly only in the postscript version, 
not in the pdf version downloaded from arxiv.org]}
 
}
\label{MCfigure.fig}
\end{figure}

Without a good estimate of age and metallicity, it is difficult to use the
light of a cluster to constrain the underlying stellar models.
Methods that explicitly deal with stochastic fluctuations when 
estimating the properties (mass, age, metallicity)
of small or intermediate mass clusters from
their integrated fluxes are in their infancy. 
The approach we are currently developing is a Bayesian one, 
which aims at determining the age, metallicity and mass probability 
distribution of observed clusters, based on their broad band fluxes.
Mass (or an analog of mass, such as the total number of stars above 
0.1\,M$_{\odot}$) is a variable that 
comes into the problem because (i) the
colour probability distributions of clusters of a given age and
metallicity depend strongly on mass, and (ii) even if dynamical
masses are available, the uncertainties about the lower end of the 
IMF make the translation into the relevant average number of 
bright stars uncertain. 

We have started a campaign of Monte-Carlo (MC) simulations, that we will
use as a database for establishing the joint probability distributions of 
mass, metallicity and age of clusters with given photometric properties
(Fouesneau et al., in preparation).
A first simple example is given in Figure \ref{MCfigure.fig}. 
It is seen that even when 
five absolute fluxes are available across the spectrum, the range of 
possible ages remains broad when dealing with clusters of $10^3$ stars.
Varying the assumed number of stars changes the results significantly,
even when only colours and no absolute fluxes are used. We are in the 
process of exploring the effects of metallicity and extinction.

The simulations are currently being extended. Clearly, the final
age probability distribution obtained for an individual cluster will depend 
on the distribution of ages, metallicities and total masses in the MC database: 
these are the priors of the Bayesian approach. The condition that
colours should be within a rigid error box (as in Fig. \ref{MCfigure.fig})
can be replaced with a continuous weighting scheme under the 
assumption that observational errors are gaussian. Globally, the method 
we are developing is a close analog of the one introduced by 
Kauffmann et al. (2003) for the study of star formation histories in the 
Sloan Digital Sky Survey galaxies, except that the variety of observable 
properties has completely different origins in both contexts.

The above approach is initially designed for the study of populations of
star clusters in the context of a host galaxy. What can we learn here
about red supergiants? Our current numerical explorations leave us with
the impression that it will be difficult to constrain red supergiant
theory with clusters of 10$^3$ solar masses, while it should become possible
with several 10$^4$ solar masses. In order to make us reject a stellar model, 
the colours of a cluster have to lie in a region of the
multi-dimensional colour-magnitude diagram that is not populated
(significantly) in spite of the stochastic fluctuations. 
The comparison of synthetic
colour-magnitude diagrams constructed with MC simulations based on 
different stellar assumptions shall tell us which colours and fluxes
are most discriminant. 

As an alternative approach, one might attempt to fit the colours or spectrum
of a low mass cluster with a combination of (i) the synthetic spectrum
of a well populated lower IMF (up to near the turn-off) 
and (ii) the spectra of a library of luminous stars. 
In some cases, it should be possible to obtain strong 
constraints on the individual nature of the handful of most luminous stars.
However, this will work best when the red supergiants are very few, in which
case the cluster as a whole will not be much brighter than a single
red supergiant star. Stochasticity will be important even for stars
near the turn-off, and the constraints on stellar evolution will 
probably not be much more stringent than those obtained from field stars in
the host galaxy. It would be worth testing these statements with 
simulations.

\section{Conclusion}

Red supergiants are responsible for large stochastic fluctuations in
the integrated energy distribution of all but the most massive 
young star clusters. These massive clusters provide useful 
tests for stellar evolution models, and should be exploited using
high signal-to-noise, co-spatial spectra covering UV to optical
wavelengths. New tools that deal with stochastic fluctuations explicitely
need to be developed further, if clusters of more normal masses are
to provide robust constraints on red supergiant models. We hope to report
progress in this area soon.


\begin{thebibliography}{}
\bibitem[Barbaro \& Bertelli (1977)]{BB77}
  Barbaro, C. \& Bertelli, C. 1977, A\&A 54, 243
\bibitem[Bastian et al. (2007)]{B_M82F_07}
  Bastian, N., Konstantopoulos, I., Smith, L.J., et al. 2007, MNRAS 379, 1333
\bibitem[Bressan et al. (1993)]{Bress93}
  Bressan, A., Fagotto, F., Bertellli, G. \& Chiosi, C. 1993, A\&AS 100, 647
\bibitem[Bruzual (2001)]{Bruzual01}
   Bruzual, G.A. 2001, in Extragalactic Star Clusters, IAU Symp. Ser. 207, 616 
\bibitem[Cervi\~no et al. (2002)]{Cervino02}
  Cervi\~no, M., Valls-Gabaud, D., Luridiana, V., Mas-Hesse, J.M. 2002,
  A\&A 381, 51
\bibitem[Cervi\~no et al. (2006)]{CervLuri06}
     Cervi\~no, M., \& Luridiana, V. 2006, A\&A 451, 475
\bibitem[Decressin et al. (2007)]{Decr07}
    Decressin, T., Charbonnel, C., Meynet, G. 2007, A\&A 475, 859
\bibitem[D'Ercole et al. (2008)]{DErcol08}
   D'Ercole, A., Vesperini, E., D'Antona, F. et al. 2008, MNRAS 391, 825
\bibitem[Fioc \& Rocca-Volmerange (1997)]{FRV97}
   Fioc, M. \& Rocca-Volmerange, B. 1997, A\&A 326, 950
\bibitem[Gallagher \& Smith (2007)]{GS07}
  Gallagher, J.S., III \& Smith, L.J. 2007, in Why Galaxies Care About 
  AGB Stars, Ed. F. Kerschbaum, C. Charbonnel, R.F. Wing, ASP Conf. Ser. 378, 
  370
\bibitem[Girardi \& Bica (1993)]{GB93}
  Girardi, L., \& Bica, E. 1993, A\&A 274, 279
\bibitem[Girardi et al. (2000)]{G2000}
  Girardi, L., Bressan, A., Bertelli, G., Chiosi, C. 2000, A\&AS 131, 371
\bibitem[Gonz\'alez Delgado et al. 2005]{GDetal05}
    Gonz\'alez Delgado, R.M., Cervi\~no, M., Martins, L.P.,
    Leitherer, C., Hauschildt, P.H. 2005, MNRAS 357, 945
\bibitem[Kauffmann et al. (2003)]{Kauffm03}
   Kauffmann, G., Heckmann, T.M., White, S.D.M. et al. 2003, 
   MNRAS 341, 33
\bibitem[Lan\c{c}on \& Mouhcine (2000)]{LanconMouh00}
     Lan\c{c}on, A., \& Mouhcine, M. 2000, in Massive Stellar Clusters,
     ed. A. Lan\c{c}on, \& C.M. Boily, ASP Conf. Ser. 211, 34
     (corrected equation: arXiv:astro-ph/0003451v2)
\bibitem[Lan\c{c}on et al. (2007)]{LHLM07}
     Lan\c{c}on, A., Hauschildt, P., Ladjal, D. \& Mouhcine, M., 2007,
     A\&A, 468, 205 
\bibitem[Lan\c{c}on et al. (2008)]{Letal08}
     Lan\c{c}on, A., Gallagher, J.S., III, Mouhcine, M., Smith, L.J.,
     Ladjal, D. \& de Grijs, R. 2008, A\&A 486, 165
\bibitem[Meynet \& Maeder (2005)]{MeyMae05}
   Meynet, G. \& Maeder, A. 2005, A\&A 429, 581
\bibitem[Schaller et al. (1992)]{Schaller92}
     Schaller, G., Schaerer, D., Meynet, G., Maeder, A. 1992, A\&AS 96, 269
\bibitem[V\'azquez et al. (2007)]{Vetal07}
     V\'azquez, G.A., Leitherer, C., Schaerer, D., Meynet, G., Maeder, A.
     2007, ApJ 663, 995
\end{thebibliography}
\end{document}